\documentclass[conference]{IEEEtran}

\usepackage{cite}
\usepackage{color}
\usepackage{multirow}
\usepackage{booktabs,multirow,array}
\usepackage[table]{xcolor} 
\usepackage{ltxcmds}
\usepackage{footnote}
\usepackage{amsthm}

\usepackage[table]{xcolor}
\usepackage{subfigure}
\usepackage{lipsum}
\usepackage{amssymb} 
\usepackage{pdflscape}
\usepackage{afterpage}

\usepackage{pifont}
\newcommand{\xmark}{\ding{55}}%
\ifCLASSINFOpdf
\usepackage[pdftex]{graphicx}
\else
  \usepackage[dvips]{graphicx}
\fi

\begin{document}
\title{A Conceptual Trust Management Framework under Uncertainty for Smart Vehicular Networks}

\author{\IEEEauthorblockN{Vishal Venkatraman$^{\ast}$, Shantanu Pal$^{\ast}$, Zahra Jadidi$^{\ast\mathsection}$, Alireza Jolfaei$^{\ast\ast}$}
\IEEEauthorblockA{$^{\ast}$School of Computer Science, Queensland University of Technology, Brisbane, QLD 4000, Australia\\ 
{$^{\ast\mathsection}$School of Information and Communication Technology, Griffith University, Gold Coast Campus, QLD 4222, Australia}\\
{$^{\ast\ast}$School of Computer Science, Macquarie University, Sydney, NSW 2109, Australia}\\
{venkatramanvishal.pinayour@connect.qut.edu.au,
(shantanu.pal,
zahra.jadidi)@qut.edu.au,
alireza.jolfaei@mq.edu.au} {}}}




\maketitle


\begin{abstract}
Trust is a fundamental concept in large-scale distributed systems like the Internet of Things (IoT). Trust helps to resolve choices into a decision. However, the trust calculation depends on the amount of uncertainty present in data sources. Trust in an IoT network is proportional to the amount of uncertainty generated by such sources as hardware malfunctions, network stability, adversarial issues, and the nature of data exchanged between the entities. The relationship between trust and uncertainty warrants approaches designed to maximize the former quality whilst minimizing the latter. Unfortunately, there is no consensus on an approach to ensure the trustworthiness of IoT networks, in particular, addressing the uncertainty issues in a fine-grained way. This paper aims to explore a generalized framework designed to manage trust in IoT networks of varying scales. In the proposed framework, several sources of uncertainty are expressed as quantities, trust ratings are calculated for individual entities in an IoT network, and a network model capable of effectively distributing workloads to trustworthy nodes is proposed. We consider a practical use case of smart vehicular networks. By realizing this paper, a standardized approach to building trustworthy IoT networks can be established, which can further guide subsequent works in the field of trust management under uncertainty.
\end{abstract}


\IEEEpeerreviewmaketitle

\section{Introduction}
\label{introduction}
Internet of Things (IoT) applications and services are increasing in everyday living spaces, public and private utilities, and workplaces alike due to their potential for enabling systematic problem-solving abilities by automating and managing repetitive tasks in a faster way~\cite{ref8}. The prevalence of such networks has been made possible by a contemporary movement in society that emphasizes a pervasive inter-connectivity of smart electronic devices (also referred to as \textit{things}) through the Internet \cite{rabehaja2019design}. In the IoT, users and devices must interact with one another, often in uncertain circumstances, which leads to uncertainty in the system. Sharing information under such uncertainty further introduces security, privacy and trust challenges due to the nature of the IoT systems (e.g., resource limitation, network and device heterogeneity, etc) \cite{pal2018fine} \cite{pal2018policy}.

The concept of uncertainty comprises multiple perspectives and interests in different scientific disciplines. The impact of uncertainty is significant to the performance of a system. If the outcomes of a process are known (i.e., in a certain state), there is a high confidence level for decision making. Likewise, the confidence levels decrease with increasing uncertainty in the outcomes. The certain and uncertain states denote the degree of uncertainty (i.e., the quality and the usefulness of information) present in the system, and the value of confidence can help to measure a system's performance~\cite{frederiksen2014trust}. 

\begin{figure}[t]
 \centering
    \includegraphics[scale=.5]{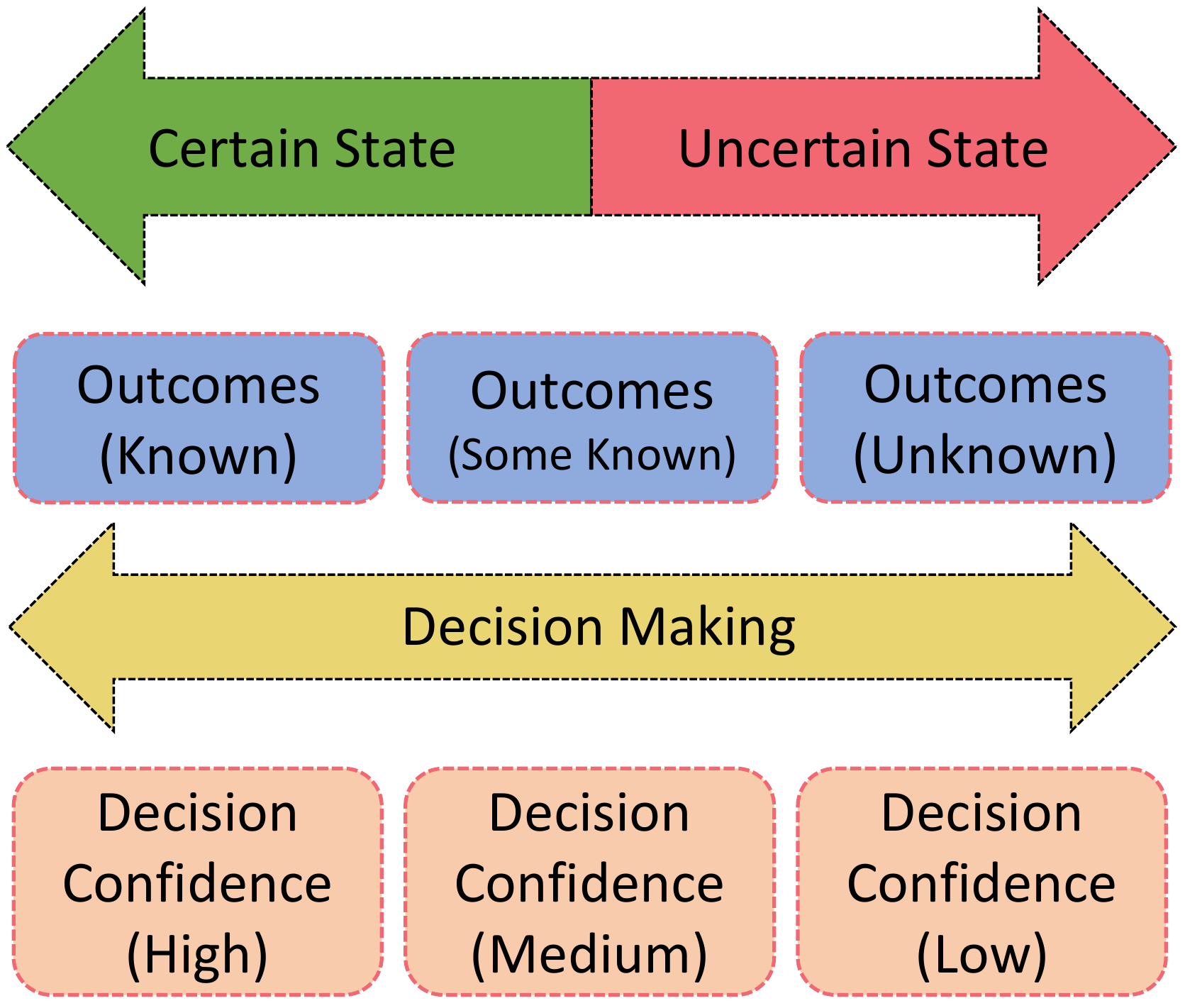}
    \caption{Impact of decision making in uncertain and certain states.}
    \label{uncertain-view}
    \small
\end{figure}

Trust can be represented as a subjective belief of an entity to another in a particular context. Trust assists in resolving choices into decisions~\cite{sicari2015security}. These are coupled with the notion of uncertainty. Therefore, to allow for decision making, it is crucial to consider two issues, (i) a representation of uncertainty, and (ii) a representation describing the impact of uncertainty associated with a particular task. In Fig.~\ref{uncertain-view}, we illustrate a simplistic association of uncertainty with decision making.

Consequently, diverse research has been conducted to examine the sources and implications of uncertainties in various multidisciplinary areas, including computing and social sciences. However, a study on uncertainty containing a detailed discussion of sources and types of uncertainty to see its impact on smart vehicular networks is lacking in the present literature. In this paper, we intend to observe uncertainty from the confidence of outcomes that determine information quality (in a decision) and look deeply into the impact of uncertainty in smart vehicular networks as a use case. Note, we refer to smart vehicular networks as IoT-enabled smart systems. We present different uncertainties and their various sources considering an IoT system. We further show how these various sources of uncertainty propagate in an IoT system and their associations with trust-related issues in smart vehicular networks. Note, we consider two well-known uncertainty types, namely aleatoric (i.e., uncertainty arising from noise or randomness) and epistemic (i.e., uncertainty arises from a lack of knowledge in systems) uncertainty~\cite{magruk2016uncertainty}.

With the increased automation based on IoT, the underlying uncertainty reduces utility and increases automation risks to a greater extent \cite{schatz2019security} \cite{pal2021analysis}. The risks are even crucial given the particular nature of the system, e.g., resource limitation, device portability, heterogeneity in networks, and high mobility of the system. For example, consider a smart vehicular network that deals with a massive number of connected vehicles, traffic management systems, and cross-domain authentication issues to provide a more efficient transportation facility. In such a system, failure to identify and comprehend uncertainty can significantly impact the system's performance~\cite{jeong2021comprehensive}. 

The motivation of this study is to couple uncertainty management with the notion of smart vehicular networks as a use case. This encompasses the significance of mining, integration, and aggregation of uncertain data coming from different sources. We also introduce a conceptual framework that serves as a reference for future attempts at constructing secure, reliable, and trustworthy IoT networks. The contributions of the paper can be summarized as follows:  

\begin{itemize}
    \item We provide a detailed quantification or qualification of various uncertainties derived from the different sources.
    \item We design a conceptual trust management framework to measure the trustworthiness of individual nodes in a large-scale IoT network (e.g., smart vehicular networks).
    \item Our study shows that the proposed design allows for distributed information sharing in a trustworthy manner under uncertainty.
\end{itemize}


The rest of the paper is organized as follows. Section \ref{background} presents the background and related work on trust and uncertainty. In Section~\ref{proposedframework}, we introduce the proposed framework. Section~\ref{resultsandevaluation}, provides an evaluation of the framework. Finally, in Section~\ref{conclusion}, we conclude the paper with future work.

\section{Background and Related Work}
\label{background}
In this section, we present a short background and related work on trust and uncertainty.

\subsection{Defining Uncertainty}
Uncertainty can be seen as a lack of quantifiable knowledge about some occurrences~\cite{ref14}. The presence of uncertainty indicates deviation(s) in the expected state(s) of a system. 
Uncertainty is also seen as a lack of the necessary information to make decisions in a system. It leads to the decision-maker synthesizing the necessary information using subjective and relatively limited evaluations. Furthermore, uncertainty can occur when it is not possible to foresee the complete set of consequences of acting. It leads to a decision-maker potentially deciding based on incomplete or unsound information~\cite{ref6}. We consider uncertainty as a probability distribution for further computation in the model.

\subsection{Sources of Uncertainty}
As stated earlier, in this paper, we consider two types of uncertainty, namely, \textit{aleatoric} and \textit{epistemic}. In Table~\ref{tab:common}, we list the various sources of uncertainties. We observe that the most frequent uncertainty source involves the hardware limitations of the devices. Other notable sources of uncertainty include the specific data management processes, network design and network stability, as well as device heterogeneity. 

\subsection{Uncertainty and Notion of Trust}
Trust can be seen as an extent of confidence with which an entity can ensure to other entities specific services tailored for given contexts and quality~\cite{ref19}. In the context of an IoT system, the ideas of trust and uncertainty are linked by their proportional nature. That said, overall trust in a system can be maximized with minimized uncertainty.

\begin{table*}[t]
\small
\centering
\caption{Common sources of uncertainties listed in the existing literature.}
\label{tab:common}

\scalebox{1}{
\begin{tabular}{l*{8}{c}r}
\hline  

Uncertainty & Magruk et~al. & Tissaoui et~al. & Kiureghian & Maddar et~al. & Abera et~al. & Ismail et~al. & Hussain et~al. \\[0.5ex] 
Sources & \cite{ref17} & \cite{ref26} & \cite{ref13} & \cite{ref16} & \cite{ref1} & \cite{ref11} & \cite{ref10} \\ [0.5ex] 
\hline

Devices Heterogeneity & \checkmark & \checkmark & \xmark & \xmark & \xmark & \checkmark & \xmark \\ [0.5ex]

Data Quality & \xmark & \checkmark & \xmark & \xmark & \checkmark & \xmark & \checkmark \\ [0.5ex] 

Data Management & \checkmark & \xmark & \xmark & \checkmark & \xmark & \checkmark  & \checkmark  \\ [0.5ex] 


Network Design & \xmark & \checkmark & \checkmark & \xmark & \xmark & \checkmark & \checkmark \\ [0.5ex] 

Network Scalability & \checkmark & \checkmark & \xmark & \checkmark & \xmark & \xmark & \xmark \\ [0.5ex] 

Network Stability & \xmark & \xmark & \checkmark & \checkmark & \xmark & \checkmark & \checkmark \\ [0.5ex] 

Privacy Protection & \checkmark & \checkmark & \xmark & \xmark & \checkmark & \xmark & \xmark \\ [0.5ex] 

Hardware Malfunctions & \checkmark & \checkmark & \checkmark & \checkmark & \checkmark & \checkmark & \checkmark \\ [0.5ex] 

Quality of Service & \xmark & \checkmark &  \xmark & \xmark & \checkmark & \xmark & \xmark \\ [0.5ex] 

Geographical Dispersal & \xmark & \xmark & \checkmark & \xmark & \xmark & \checkmark & \xmark \\ [0.5ex] 

Environmental Effects & \xmark & \xmark & \xmark & \checkmark & \xmark & \checkmark & \xmark \\ [0.5ex] 

\hline
\end{tabular}
}
\end{table*}

\subsection{Uncertainty and Trust for Smart Vehicular Networks}


There are several works that consider uncertainty and trust-building together for dynamic systems \cite{siddiqui2021survey}. In addition, many other works propose a trust management framework for smart vehicular networks. But the issue of trust management under uncertainty when considering the smart vehicular networks are lacking in the present literature. For instance, proposals \cite{ref5} and \cite{ref7} discuss uncertainty issues in cyber physical systems (CPS). In these proposals, uncertainties are identified on the application, infrastructure, and integration levels of the CPS. Unlike our motivation, they do not discuss trust management under uncertainty. 
Proposal \cite{ref17} discusses the concept of uncertainty in a practical context of smart buildings. Uncertainties are considered using subjective logic. Proposal \cite{ref26} discusses the challenges of uncertainties in an IoT system. It emphasises the impact of uncertainty on the quality of data transmission in the IoT pipeline. But these proposals do not consider how uncertainty impacts in distributed trust building.

Fernandez-Gago et al. \cite{ref9} design a trust management framework for an IoT network that relies on trust, privacy, and identity requirements. Maddar et al. \cite{ref16} offer an intrusion detection model for an IoT network composed of wireless sensor networks (WSN). Abera et al. \cite{ref1} employ the process of remote attestation, to monitor sparsely distributed nodes in an IoT network. Pal et al. \cite{ref20} propose a trust management system for large-scale IoT networks focused on improving access control mechanisms for resource-constrained edge devices. Ruan et al. \cite{ref23} discuss a general trust management framework for IoT networks. The proposed trust management model utilizes the overall trustworthiness of a node and the measure of confidence in its own evaluation of trustworthiness as metrics to manage trust in the system. While these proposals discuss the notion of trust in IoT systems, they do not consider the impact of uncertainty within the model. 

Proposal \cite{garcia2018security} presents a trust management framework for smart vehicular networks. In this framework, vehicles can detect a compromised vehicle (e.g., attacked by a malicious agent for performing malicious activities) in proximity and ignore communications with them. In \cite{zhang2018machine}, a machine learning approach is used to build trust management for vehicular networks. In this approach, a trust model is devised based on the behaviour of nodes located in proximity for forwarding packets. Other proposals, e.g.,  \cite{huang2017vehicular}, \cite{padmapriya2022secure}, and \cite{ayobi2020lightweight}, use fog computing and blockchain-based solutions for trust management in smart vehicular networks. However, these proposals do not consider the uncertainty issue in the model during trust-building. Therefore, no organized and shared approach lists the various implications of uncertainty and how it can 
impact the overall performance of an IoT-enabled system, e.g., a smart vehicular network when considering distributed trust.

\section{Proposed Framework}
\label{proposedframework}
In this section, we discuss the proposed trust management framework. The objective of the framework is to: (i) determine procedures for quantifying uncertainties, and (ii) derive trust ratings from the quantities. In addition, we have designed a network model to enable a sufficiently large-scale IoT system. 

\subsection{Assumptions}
We make the following assumptions: 

\begin{itemize}
    \item \textit{Trust in the network is maximized by minimizing the impact of uncertainties}: under this assumption, the proposed framework is designed to allocate decision-making workloads to only the most trustworthy nodes in the network. 
    \item \textit{A node cannot act unless sufficiently trustworthy}: under this assumption, the proposed framework requires nodes to continuously update their trust ratings so that the system can be sufficiently trustworthy by maintaining updated information. 
\end{itemize}

It is also presumed that all nodes in the system are provisioned with the appropriate hardware and software implements to measure characteristics of uncertainties and perform decision making computations. To simplify discourse, it is assumed that each node in the network is equipped to handle the required decision making processes because discussions on specific hardware and software solutions are not within the scope of this research.

\subsection{Design Considerations}
In this section, we introduce the general composition of the proposed framework.

\begin{itemize}
    \item \textit{The framework categorizes uncertainties into aleatoric and epistemic uncertainties.}
\end{itemize}

The proposed framework considers aleatoric and epistemic uncertainties. These two categories are chosen because they can be applied to specific IoT contexts while remaining abstract enough to warrant their inclusion in a conceptual framework. It is expected that epistemic uncertainty can be expressed in qualitative terms (e.g., \textit{High}, \textit{Medium}, and \textit{Low}) while aleatoric uncertainty is expressed in quantitative terms (numerical values) \cite{ref13}.

\begin{itemize}
    \item \textit{The framework employs fuzzy logic to convert epistemic uncertainties into numerical data.}
\end{itemize}

The framework anticipates difficulties quantifying epistemic uncertainties due to the challenges in expressing the lack of knowledge in certain phenomena. To address this, we employ fuzzy logic to convert epistemic uncertainties into numerical values that can enable the trust management process. Fuzzy logic allows for the computation of linguistic descriptors like \textit{High} and \textit{Low}, which are lacking in numerical definition. Employing fuzzy logic involves the conversion of such subjective uncertainty quantities into objective numerical values through the process of fuzzification, inference and defuzzification.

\begin{itemize}
    \item \textit{The framework follows a network architecture inspired by Directed Acyclic Graphs (DAG).}
\end{itemize}

A key aspect of the proposed framework's trust management is the propagation of trust values across the network. Therefore, the framework is expected to offer nodes in the network knowledge of the extent of trustworthiness of neighbouring nodes. By publishing these trust ratings, nodes are empowered to carry out decision-making processes with only the most trustworthy nodes, thereby simultaneously distributing workloads and maximizing the trustworthiness of the result. To facilitate this behaviour, the framework follows a network architecture based on DAGs, as utilized in the IoTA cryptocurrency \cite{ref18}. 

\section{Framework Evaluation}
\label{resultsandevaluation}
The evaluation revolves around the design of the proposed framework, which is composed of procedures for trust calculation and a network model that allows for scalable distribution of workloads under uncertainty.

\subsection{Prioritized the Sources of Uncertainty}
Table~\ref{tab:prioritized} lists four types of priority and severity that we use in our framework. These sources are derived from Table~\ref{tab:common}. Sources that are cited equally in the existing literature frequently are prioritized equivalently.
\begin{table}[h]
\small
\centering
\caption{Sources of Uncertainty and Priority.}
\label{tab:prioritized}

\scalebox{1}{
\begin{tabular}{c*{1}{l}r}
\hline  
Priority/Severity & Uncertainty Sources \\ [0.5ex] 
\hline \hline
1 &	Hardware Malfunctions \\ [0.5ex] 
\hline
2 &	Data Management \\ [0.5ex] 
  &	Network Design \\ [0.5ex] 
  &	Network Stability \\ [0.5ex] 
\hline
3 &	Devices Heterogeneity\\ [0.5ex] 
  & Data Quality \\ [0.5ex] 
  &	Network Scalability \\ [0.5ex] 
  &	Privacy Protection \\ [0.5ex] 
\hline
4 &	Quality of Service \\ [0.5ex] 
  &	Geographical Dispersal \\ [0.5ex] 
  &	Environmental Effects \\ [0.5ex] 
\hline
\end{tabular}
}
\end{table}

In Table~\ref{tab:categorized_source}, the sources of uncertainty are categorized into epistemic, aleatoric or both to guide in designing the appropriate data collection procedures for each uncertainty. As mentioned earlier, it is expected that nodes in the network measure epistemic uncertainties qualitatively and aleatoric uncertainties quantitatively. 

\begin{table}[h]
\small
\centering
\caption{Categorized Sources of Uncertainties Based on Types.}
\label{tab:categorized_source}

\scalebox{1}{
\begin{tabular}{l*{1}{l}r}
\hline  
Category (Epistemic/Aleatoric/Both) & Uncertainty Sources \\ [0.5ex] 
\hline \hline
Both&	Hardware Malfunctions \\ [0.5ex] 
Aleatoric&	Data Management \\ [0.5ex] 
Both&	Network Design \\ [0.5ex] 
Both&	Network Stability \\ [0.5ex] 
Both&	Devices Heterogeneity \\ [0.5ex] 
Aleatoric&	Data Quality \\ [0.5ex] 
Epistemic&	Network Scalability \\ [0.5ex] 
Both&	Privacy Protection \\ [0.5ex] 
Aleatoric&	Quality of Service \\ [0.5ex] 
Aleatoric&	Geographical Dispersal  \\ [0.5ex] 
Both&	Environmental Effects  \\ [0.5ex] 
\hline
\end{tabular}
}
\end{table}

\subsection{Calculation of Trust Rating}
\begin{figure}[ht]
    \centering
    \includegraphics[scale=1.25]{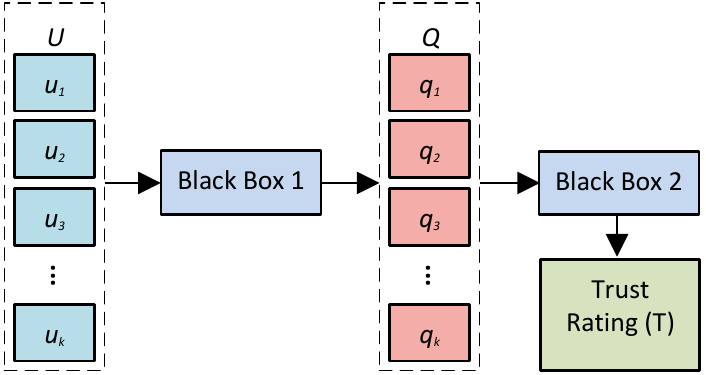}
    \caption{An outline of the computation of trust rating.}
    \label{fig-1}
\end{figure}

Fig.~\ref{fig-1} illustrates the process of transforming uncertain data obtained from nodes to a trust rating for the node. 

Once a list of uncertainties and the means to measure them have been identified, the framework defines each uncertainty as a variable \textit{u\textsubscript{i}} such that each uncertainty is a part of the set \textit{U} of size \textit{n}. This is expressed as follows, 

\setlength{\belowdisplayskip}{0pt} \setlength{\belowdisplayshortskip}{0pt}
\setlength{\abovedisplayskip}{0pt} \setlength{\abovedisplayshortskip}{0pt}
\begin{equation}
U=\left\{u_1,u_2,u_3,\dots,u_n \right\}
\end{equation}
\begin{equation}
|U|=n
\end{equation}

Given the existence of both aleatoric and epistemic uncertainties, it is expected that \textit{U} will be the union of the subsets \textit{U\textsubscript{A}} and \textit{U\textsubscript{E}} defined below, where \textit{U\textsubscript{A}} represents the subset of all aleatoric uncertainties and \textit{U\textsubscript{E}} represents the subset of all epistemic uncertainties.

\begin{equation}
U_A=\left\{u_{A1},u_{A2},u_{A3},\dots,u_{Aj} \right\}
\end{equation}
\begin{equation}
U_E=\left\{u_{E1},u_{E2},u_{E3},\dots,u_{Ek} \right\}
\end{equation}

\textit{U\textsubscript{A}} has \textit{j} elements and \textit{U\textsubscript{E}} has \textit{k} elements, and the sum of elements in both sets should be equivalent to the number of elements in set \textit{U}. In other words, the framework considers all possible uncertainties to be either aleatoric or epistemic. 

\begin{equation}
|U_A |=j
\end{equation}
\begin{equation}
|U_E |=k
\end{equation}
\begin{equation}
U_A\cup U_E=U
\end{equation}

Since Table~\ref{tab:categorized_source} offers sources of uncertainty that can be considered both aleatoric and epistemic in nature, it is expected that adopters of this framework decompose the provided sources of uncertainty in a manner that limits the categorization of individual uncertainties to either the aleatoric or epistemic types.  

The proposed framework computes aleatoric and epistemic uncertainties using different approaches, as outlined earlier. The complete set \textit{U} is the input required by Black Box 1, which is represented by \textit{B\textsubscript{1}(U)}, and is expected to output a set \textit{Q}. A discussion of Black Box 1 and 2 are given below. 

\subsubsection{Black Box 1}

It is responsible for taking a set of uncertainties \textit{U} and quantifying or approximating them appropriately, thereby offering an output of \textit{Q}, which is the set of numerical uncertainties with \textit{n} elements.

\begin{equation}
Q=\left\{q_1,q_2,q_3,\dots,q_n\right\}
\end{equation}
\begin{equation}
|Q|=n
\end{equation}

Each element \textit{q\textsubscript{i}} in \textit{Q} is derived by applying Black Box 1 to the corresponding variable \textit{u\textsubscript{i}} in \textit{U}. Thus, 
\begin{equation}
q_i=B_1 (u_i)
\end{equation}
Similar to \textit{U}, \textit{Q} is the union of the aleatoric and epistemic subsets \textit{Q\textsubscript{A}} and \textit{Q\textsubscript{E}}. \textit{Q\textsubscript{A}} is of size \textit{j} and \textit{Q\textsubscript{E}} is of size \textit{k}. Thus,
\begin{equation}
Q_A\cup Q_E=Q
\end{equation}
\begin{equation}
|Q_A |=j
\end{equation}
\begin{equation}
|Q_E |=k
\end{equation}

\textit{Q\textsubscript{A}} is subjected to `Monte Carlo' experiments, which utilize random sampling methods to obtain numerical values. This approach was used by \cite{ref4} to quantify the reliability of IoT networks. Using the input \textit{u\textsubscript{Ai}}, Black Box 1 runs a simulation to estimate the extent of uncertainty represented by the input. 
\begin{equation}
B_1 (U_A )=Q_A  
\end{equation}

The corresponding output \textit{q\textsubscript{Ai}} is a numerical quantity of aleatoric uncertainty, and the resultant set \textit{Q\textsubscript{A}} can be processed further by Black Box 2 to obtain the trust rating for a node.
\textit{Q\textsubscript{E}}, meanwhile, is subjected to the `Mamdani Fuzzy Inference System' used by \cite{ref12}, which is a fuzzy logic system consisting of the following three stages \cite{ref12}. 
\begin{itemize}
	\item Fuzzification of the input \textit{u\textsubscript{Ei}}, which involves converting the input into linguistic fuzzy logic variables, e.g., \textit{High}, \textit{Medium}, and \textit{Low}. 
	\item Fuzzy Inferencing, which is the process of operating on linguistic variables using a set of fuzzy rules. 
	\item Defuzzification, which is the process of converting the inferred results into a numerical output \textit{q\textsubscript{Ei}}.
\end{itemize}

\begin{figure}[ht]
    \centering
    \includegraphics[scale=1.1]{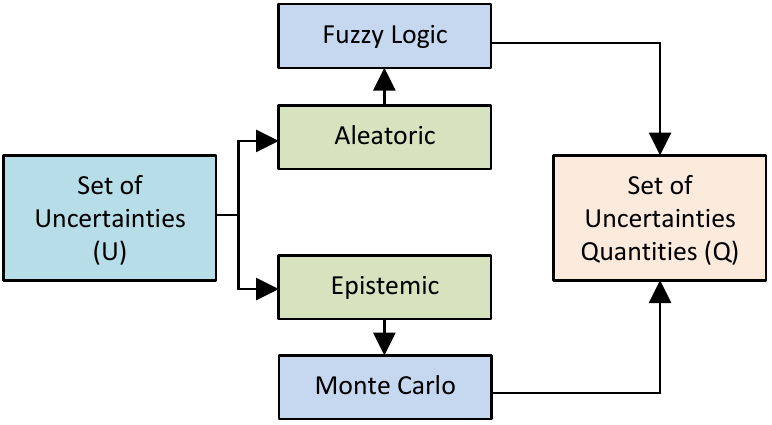}
    \caption{Black Box 1 - Quantification of uncertainties.}
    \label{fig-2}
\end{figure}

Using the input \textit{u\textsubscript{Ei}}, Black Box 1 translates non-numerical descriptors to numerical values. 
\begin{equation}
B_1 (U_E )=Q_E  
\end{equation}
The output \textit{q\textsubscript{Ei}} is a numerical quantity of epistemic uncertainty, and the resultant set \textit{Q\textsubscript{E}} can be processed further by Black Box 2 to obtain the required trust rating for a node. Fig.~\ref{fig-2} summarizes the logic flow of Black Box 1. 

\subsubsection{Black Box 2}
To evaluate a trust rating, it is necessary to apply weights to the set \textit{Q}, as Table~\ref{tab:prioritized} prioritizes certain sources of uncertainty over others. Thus, the framework expects a set of weights \textit{W} of size \textit{n} to be defined \cite{ref5}. 
\begin{equation}
W=\left\{w_1,w_2,w_3,\dots,w_n\right\}
\end{equation}
\begin{equation}
|W|=n
\end{equation}	
It is expected that the quantified uncertainty variable \textit{q\textsubscript{i}} will have a corresponding weight \textit{w\textsubscript{i}}. The steps to obtain the required trust rating are as follows:
\begin{itemize}
	\item Multiply variable \textit{q\textsubscript{i}} with its corresponding weight \textit{w\textsubscript{i}}. 
	\item Sum the resultant weighted uncertainty variables. 
	\item Divide the result by the sum of all weights \textit{w\textsubscript{i}} in \textit{W}. 
\end{itemize}

Equation \ref{eq-trust} summarizes the steps required to compute the trust rating \textit{T}, which is expected to exist in the range of [0, 1]. Where, 0 represents absolute distrust, and 1 represents absolute trust. 
\begin{equation}
\label{eq-trust}
\frac{\sum_{i=0}^{n}Q_i*W_i}{\sum_{i=0}^{n}W_i}=T
\end{equation}

\begin{figure}[ht]
    \centering
    \includegraphics[scale=1.25]{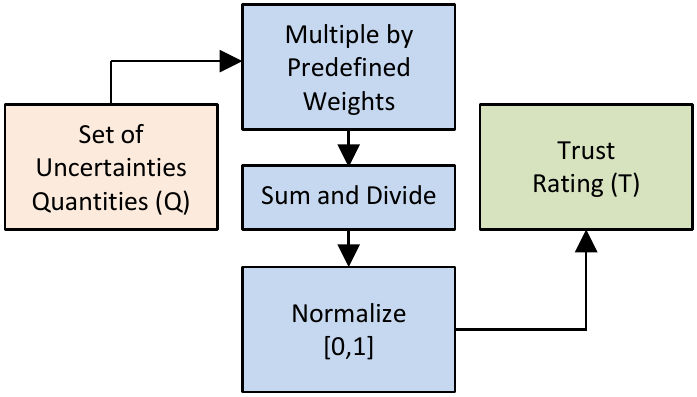}
    \caption{A simple outline of the calculation of trust ratings.}
    \label{fig-3}
\end{figure}

The value \textit{T} is the resultant trust rating of the node in question, and it will be used to determine the suitability of a node to carry out a task. A node with a value \textit{T} closer to 1 is more likely to contribute to a decision-making process than a node with a value \textit{T} closer to 0. Fig.~\ref{fig-3} summarizes the logic flow of Black Box 2. 

\subsection{Network Model}
It is necessary to define a network model that allows for distributed trust sharing and decision-making whilst being resistant to the issues associated with network scalability. Thus, it is recommended that a network model inspired by DAGs be implemented in the proposed framework. Proposal \cite{ref27} considers improving the linear blockchain-based approach with a DAG, which stabilizes network performance as the network grows. In the proposed framework, modelling the network after DAGs would satisfy two functions: firstly, it would enable the provision of a distributed trust ledger detailing the trust values of all the nodes in the network, and secondly, it would allow nodes to share the workload of decision-making with other suitably trustworthy nodes, thereby ensuring a distributed system where trustworthiness is maximized by minimizing uncertainty. 

\subsubsection{Composition of the Network}
\begin{figure}[t]
    \centering
    \includegraphics[scale=1.55]{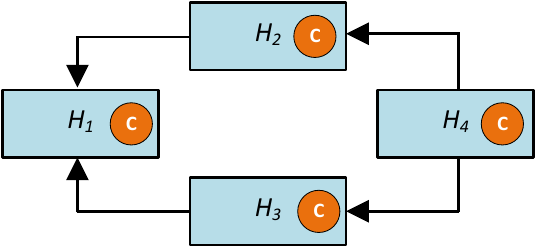}
    \caption{DAG of smaller peer-to-peer clusters.}
    \label{fig-4}
\end{figure}

In the model, nodes that are geographically close to each other are considered a part of the same cluster. Within such clusters, nodes are arranged in a peer-to-peer manner, which allows each node to communicate directly with any other node without any impositions of hierarchy. It is expected that several such clusters are merged to form a DAG of clusters. Fig.~\ref{fig-4} shows a DAG comprised of clusters \textit{H\textsubscript{1}}, \textit{H\textsubscript{2}}, \textit{H\textsubscript{3}} and \textit{H\textsubscript{4}}. 

\subsubsection{Distributed Trust Ledger}
The proposed framework maintains a Trust Ledger, which contains the trust ratings of all the nodes in the network. By using the distributed ledger approach employed by cryptocurrency systems \cite{ref28}, it is possible to host this Trust Ledger in a distributed fashion across all the clusters in the network, thereby not relying on  centralized storage space. 

When two nodes \textit{N\textsubscript{1}} and \textit{N\textsubscript{2}} within a cluster complete a round of communication with each other, \textit{N\textsubscript{1}} evaluates the trust rating of \textit{N\textsubscript{2}}, and \textit{N\textsubscript{2}} does the same for \textit{N\textsubscript{1}}. These new trust ratings are added to the Trust Ledger, where the trust rating of each node is maintained as a rolling average value. By maintaining rolling average values, the framework prioritizes nodes that have historically maintained high trust ratings over an extended period of time. 

\subsubsection{Coordinators}
When nodes are tasked with carrying out an action, the framework ensures that the workload is distributed among only the nodes with the highest trust ratings. This ensures that the amount of uncertainty in the result of the action is minimal. The distribution of workloads is the responsibility of the coordinator node: should a cluster \textit{H\textsubscript{1}} need to collaborate on a decision-making task with another cluster \textit{H\textsubscript{2}}, the coordinator nodes \textit{C\textsubscript{H\textsubscript{1}}} and \textit{C\textsubscript{H\textsubscript{2}}} act as representatives of their corresponding clusters. 

\begin{figure}[ht]
    \centering
    \includegraphics[scale=2.2]{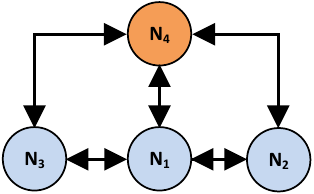}
    \caption{The peer-to-peer cluster with the node $N_4$ serving as the coordinator. }
    \label{fig-5}
\end{figure}

The coordinator node of a cluster is the node with the highest trust rating maintained over an extended period of time. Since trust ratings are updated upon the completion of each round of communication, the duties of the coordinator are assumed by different nodes over time. This approach eliminates the need for a centralized coordinator. Fig.~\ref{fig-5} displays the peer-to-peer setup of a hypothetical cluster along with its coordinator, with Table~\ref{tab:Hypothetical} displaying trust ratings of nodes in the cluster. It must be noted that the trust ratings in Table~\ref{tab:Hypothetical} were randomly generated to serve as an example. 

\begin{table}[h]
\small
\centering
\caption{Hypothetical trust ratings of nodes in a cluster.}
\label{tab:Hypothetical}

\scalebox{1}{
\begin{tabular}{c*{2}{c}r}
\hline  
Node & Trust Rating	& Coordinator \\ [0.5ex] 
\hline \hline
$N_1$	& 0.54& 	No \\ [0.5ex] 
$N_2$	& 0.79& 	No \\ [0.5ex] 
$N_3$	& 0.86& 	No \\ [0.5ex] 
$N_4$	& 0.91 & 	Yes \\ [0.5ex] 
\hline
\end{tabular}
}
\end{table}

\subsection{Discussion}
In this section, we provide some significant findings and address some points of consideration when implementing the framework given its choice of components. The proposed model is helpful for smart vehicular networks in many different aspects. The framework can gather a collection of different sources of uncertainties for distributed trust-building. It significantly helps multiple applications, including self-driving, post-accident processing, and automatic driving monitoring in a smart city environment.
Black Boxes 1 and 2 provide a computational foundation for the quantification of trust ratings but does not offer specific parameters with which to generate the required ratings. This is by design, as the framework is intended to be a conceptual one that can be extended to cater to specific contexts.



Although the network model proposed as part of the framework addresses the requirement of a distributed system and the issue of network scalability, it contributes to some issues. For instance, under the proposed framework, most of the computational burden will be on the nodes in the network. This is a consequence of distributed computing, and a solution is to ensure that the individual nodes in the system can with-stand the increased workloads. Indeed, for the framework to be successfully implemented, all nodes must be equally capable in terms of processing power. 

Since decision-making processes are to be assigned to nodes with the highest trust ratings over an extended period of time, it will take the nodes in the network an amount of time to reach `maturity' and therefore be considered trustworthy. In the meantime, the framework may not necessarily distribute workloads effectively due to a lack of adequate information on the trustworthiness of nodes. This problem has been addressed in the context of cryptocurrencies, where distributed ledger systems are supplemented by a temporary centralized coordinator that handles the distribution of data \cite{ref3} \cite{ref22}. Should such a solution be deployed, it is important that adopters of the framework be prepared to handle the risks associated with relying on a centralized authority to distribute workloads as encountered in blockchain technology. 

Moreover, the clusters in the network must contain similar numbers of nodes to enable the equal distribution of workloads. Should there be a large variance in sizes among different clusters, issues related to hardware failure and network performance may arise. This is especially important if adopters wish to extend the network rapidly in size: it is recommended that new clusters be added to the existing network, or existing clusters be expanded uniformly.

\section{Conclusion and Future Work}
\label{conclusion}
In this paper, we have proposed a framework that accomplished the challenge of ensuring trust in a distributed IoT network by identifying its sources of uncertainty and quantifying them. It offers mechanisms to monitor individual uncertainties, calculate trust ratings for nodes in the network, and a network model that enables decision-making within the network in a distributed manner. The framework can maximize the trust value in the system while minimizing the impact of uncertainties. Since the framework is meant to serve as a foundation for more context-specific implementations, adopters of the framework are encouraged to use the provided sources of uncertainty as a foundation for more rigorous definitions of uncertainty relevant to specific contexts. In future, we intend to conduct practical case studies using the proposed network model in various identified uncertainty sources to demonstrate the suitability of the proposed framework at scale.

\section*{Acknowledgment}
The authors acknowledge the support of the Commonwealth of Australia and Cybersecurity Research Centre Limited.


\ifCLASSOPTIONcaptionsoff
  \newpage
\fi

\bibliographystyle{IEEEtran}
\bibliography{bare_jrnl}

\end{document}